\documentclass[a4paper, twocolumn, reqno]{amsart}
\usepackage[letterpaper, portrait, margin=1in]{geometry}
\usepackage{graphicx, caption}
\usepackage{amsfonts}
\usepackage{amsmath,bm}
\usepackage{cite}
\usepackage{IEEEtrantools}
\usepackage{amssymb}
\usepackage{amsthm}
\usepackage{color}
\usepackage{mathtools}
\usepackage{braket}
\usepackage{enumitem}
\usepackage{multirow}
\usepackage{hyperref}
\usepackage{amsaddr}
\usepackage{balance}
\usepackage{flushend}

\theoremstyle{plain}

\title{Quantum Computing for Data Centric Engineering and Science}
\author{Steven Herbert} 
\address{Quantinuum (Cambridge Quantum), Terrington House, 13-15 Hills Rd, Cambridge, CB2 1NL,  UK \\ Department of Computer Science and Technology, University of Cambridge, UK}

\begin{document}
\sloppy

\onecolumn

    
    \begin{abstract}
In this perspective I give my answer to the question of how quantum computing will impact on data-intensive applications in engineering and science. I focus on quantum Monte Carlo integration as a likely source of (relatively) near-term quantum advantage, but also discuss some other ideas that have garnered wide-spread interest.
    \end{abstract}

\twocolumn[{%
 \centering
 \maketitle
 
 \vspace{-1cm}
}]

\section{Quantum Computing: A Very Brief Introduction}
\label{intro}

\noindent The conception of quantum computing is usually attributed to Richard Feynman, who in 1981 speculated that simulating the behaviour of a quantum mechanical system would require a computer that was itself somehow quantum mechanical in nature \cite{Feynman:1981tf,PreskillFeynman}; Manin \cite{manin} and Benioff \cite{Benioff} also espoused similar ideas at around the same time. It was David Deutsch who in 1985 then laid the groundwork for quantum computing as we now know it, by formalising a quantum mechanical model of computation, and posing well-defined mathematical problems where quantum computing offers a clear computational advantage \cite{Deutsch1985QuantumTT}. This in turn spawned a great profusion of activity in the then embryonic field of quantum computing in the late 1980s and early 1990s, leading to what remain to this day two of the crowning achievements of the field: in 1994 Peter Shor proposed a quantum algorithm for factoring in polynomial time \cite{Shor}; and in 1996 Lov Grover proposed an algorithm to search an unstructured database in time proportional to the square root of the database size \cite{Groversearch}.
\makeatletter{\renewcommand*{\@makefnmark}{}
\footnotetext{\noindent Contact: Steven.Herbert@Quantinuum.com \\ This is the arxiv version of \url{https://doi.org/10.1017/dce.2022.36} }\makeatother}

Unstructured search (in this context) is the problem where we have some $N=2^n$ elements, indexed $\{0,1\}^n$, to search through, and a `function', $f$, such that for exactly one $x \in \{0,1\}^n$, $f(x) = 1$ and $f(x) = 0$ otherwise. `Unstructured' means there is no algorithmic short-cut -- $f$ is a function in the technical sense only and does not imply it can be represented as some simple algebraic expression -- and hence classically the best (only) strategy is exhaustive search, which requires $f(x)$ to be evaluated for all $N$ elements at worse, and $N/2$ elements on average. Quantumly, we can prepare a superposition of all possible $n$-bistrings, and hence `query' $f$ for all possible $x$ in a single step, \emph{however this does not imply that quantum unstructured search completes in} $\mathcal{O}(1)$ \emph{operations}. In fact, as the answer is encoded in a quantum state it turns out that it takes at least $\mathcal{O}(\sqrt{N})$ operations to extract -- a lower bound that Grover's search algorithm achieves. This improvement from $\mathcal{O}(N)$ classical operations to $\mathcal{O}(\sqrt{N})$ quantum operations is commonly referred to as a `quadratic advantage'. 

Whilst the quadratic advantage is extremely valuable, the fact that quantum computing enables the simultaneous querying of $f$ for an exponential number of $x$ (that is, we say the problem size is `$n$' and we query $f$ for all $N=2^n$ possible $n$-bitstrings in superposition), dangles the tantalising possibility of \emph{exponential} computational advantages. To see such advantages we must move on from unstructured search to problems with some specific structure that can be attacked by quantum, but not classical algorithms. The manner in which this structure is attackable by the quantum algorithm can be a little hard to grasp, but essentially amounts to the fact that the answer we are searching for is in some sense determined by all $2^n$ queries, but in such a way that a quantum mechanical `interference' step (for which there is no analogue in classical computation) can efficiently extract the solution. This is indeed the case for Shor's factoring algorithm, where the \emph{Quantum Fourier Transform} (QFT) performs this interference step (in fact many of the most prominent proposals for super-polynomial quantum advantage use the QFT). Classically the best factoring algorithm is the number field sieve \cite{numbersieve}, which has complexity $\exp(\Theta(n^{1/3} \log^{2/3} n))$; whereas Shor's algorithm requires only $\mathcal{O}(n^2 \log n \log \log n)$ operations, where the problem size, $n$, is the number of bits required to express the number being factored. 

For the past 25 years, Shor's and Grover's algorithms have been mighty pillars upon which many other proposals for quantum algorithms have been built, and the computational complexity thereof continues to provide some insight into the sorts of advantage we should expect from quantum algorithms: if the algorithm is tackling a task with little structure, then we expect a quadratic (or other polynomial) advantage; whereas if there is structure that can be exploited by a quantum interference step (such as the QFT) then we can get a super-polynomial speed-up. (Scott Aaronson recently posted a very nice and concise article about the role of structure in quantum speed-ups \cite{AaronsonStructure}.)

\begin{figure*}[t]
\begin{center}
\includegraphics[width=0.9\textwidth]{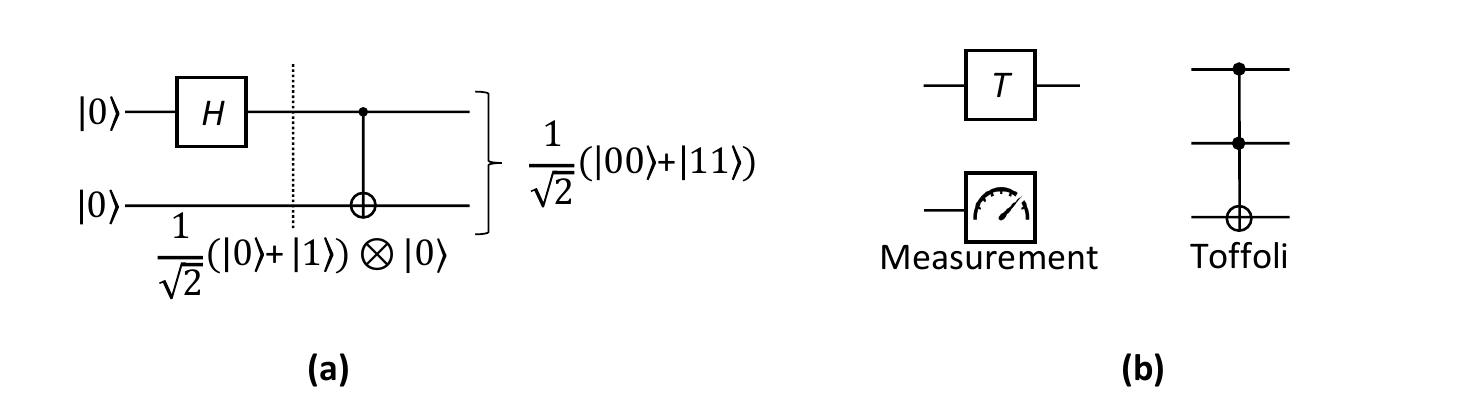}
\captionsetup{width=.9\linewidth}
\caption{The quantum algorithms discussed in this paper are generally formulated in terms of the quantum circuit (or quantum gate) model. Each wire corresponds to a qubit and the gates are unitary matrices, the qubits are initialised in a computational basis state $\ket{0} = [1,0]^T$ or $\ket{1} = [0,1]^T$, which are analogous to the 0 and 1 states of classical bits. Qubits differ as they may be put in a superposition of the computational basis states, for example, the Hadamard ($H$) gate in the circuit in \textbf{(a)} puts the first qubit in the superposition $(1/\sqrt{2})(\ket{0} + \ket{1})$ (quantum states are such that the squared moduli of the coefficients of the computational basis states sum to one). The qubits in the circuit are composed using the tensor product, and hence the full state after the Hadamard gate is $(1/\sqrt{2})(\ket{0} + \ket{1})\otimes \ket{0}$ (which can equivalently be expressed $(1/\sqrt{2})(\ket{00} + \ket{10})$). The next gate is the two-qubit CNOT, which transforms the state into the `\textit{Bell state}' $\ket{\Phi^+} = (1/\sqrt{2})(\ket{00} + \ket{11})$ — a state which cannot be expressed as a tensor product of two single-qubit states, and is thus referred to as an \textit{entangled} state. As well as the Hadamard and CNOT, some further important gates are shown in \textbf{(b)}: the $T$ gate, Toffoli and measurement. Together $H$, $T$ and CNOT form a universal gate set — any quantum circuit can be expressed (to arbitrary precision) as a circuit containing just these; however, the Toffoli gate is also a useful primitive as it implements the classically universal NAND gate as a (three qubit) unitary operation. Measurements are needed to extract information from the quantum state, and a (single-quit computational basis) measurement yields a single classical bit. The measurement outcome is random, and each computational basis state is measured with probability equal to its coefficient's modulus squared, for example if the state $(1/\sqrt{2})(\ket{0} + \ket{1})$ is measured, then the classical bits 0 and 1 are each measured with 50\% probability, this is known as the `Born rule'}
\label{f0}
\end{center}
\end{figure*}

A further, important point to note is that all of the `canonical' quantum algorithms presume an abstract model of quantum computation, which is innately noiseless (quantum noise occurs when the environment randomly perturbs the quantum state such that it departs from that predicted by the abstract model of quantum computation). It was therefore a substantial and highly important breakthrough when it was shown that real, noisy, quantum hardware can efficiently simulate the noiseless model of quantum computation in principle owing to the celebrated \textit{threshold theorem} \cite{thresh0, thresh1, thresh2, thresh3}. However, in practice, this still requires a quantum error-correction overhead that takes the noiseless model out of reach of near-term quantum hardware. In the past several years significant attention has been given to the question of whether useful quantum advantage can be obtained by computing with noisy qubits, that is, without quantum error-correction. For this setting, John Preskill coined the term `NISQ' (\textit{noisy intermediate-scale quantum [computer]}) \cite{Preskill2018}, and it has become commonplace to speak of the `NISQ-era' (computing with noisy qubits) which will eventually give way to the `full-scale era' (when quantum error correction will mean that we can essentially treat the qubits as noiseless), although I shall later argue that this is something of a false dichotomy.

In general, both `NISQ' and `full-scale' quantum algorithms are usually formulated using the \textit{quantum circuit model}\footnote{Alternatives to the quantum circuit model include \textit{linear optical quantum computing} \cite{Knill2001ASF}, \textit{adiabatic quantum computation} \cite{adiabatic} and \textit{measurement-based quantum computation} \cite{MBQC}.}, which is briefly introducted in Fig.~\ref{f0}. It is common to use circuit \textit{depth} (the number of layers of operations) as a proxy for computational complexity and in the case of NISQ algorithms, the circuit depth dictates the number of operations that must be performed with the state remaining coherent (i.e., before the noise becomes too great and the information contained within the state is lost). So it follows that, when designing quantum algorithms with resource constraints in mind, it is important to keep the circuit depth as low as possible -- in order to achieve the computation within the physical qubit coherence time (NISQ) or with as little error correction as possible (full-scale).

\begin{figure}[!t]
\begin{center}
\includegraphics[width=0.9\linewidth]{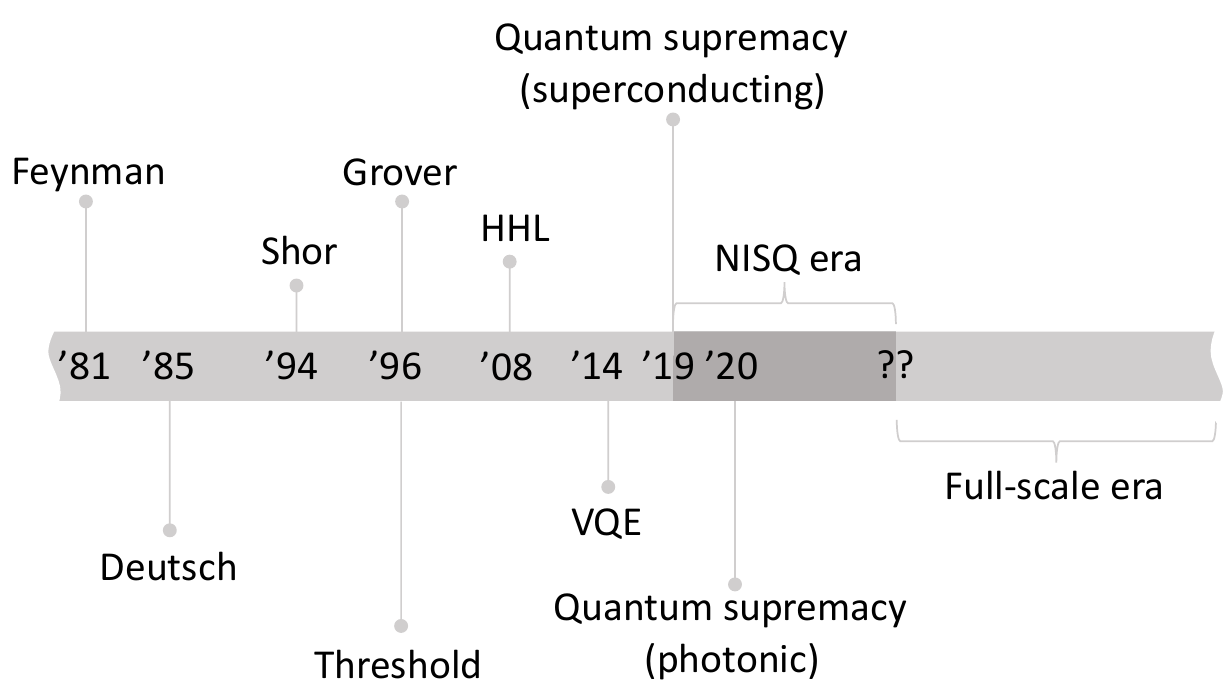}
\end{center}
\captionsetup{width=.9\linewidth}
\caption{A timeline of some of the most important results in quantum computing. The variational quantum eigensolver (VQE) \cite{vqe} is widely acknowledged as one of the most promising NISQ algorithms; and Google's demonstration of (superconducting) quantum supremacy may be taken as the start of the NISQ-era (in the sense that this was the first demonstration of a quantum algorithm significantly outperforming its classical counterpart). Quantum supremacy on a photonic quantum computer was first claimed by Jian-Wei Pan's group \cite{supremephoton}, and later by Xanadu \cite{supremephoton2}
}
\label{f1}
\end{figure}

Fig.~\ref{f1} summarises some of the most important breakthroughs in quantum computing, and for further information the reader is directed to \textit{Nielsen and Chuang} \cite{nielsenchuang2010} which remains the authoritative textbook on the subject. The \textit{quantum algorithm zoo} \cite{qaz} also provides a catalogue of many suggested quantum algorithms -- although the total number of algorithms can be somewhat misleading: many of the listed algorithms amount to different instances and applications of the same essential quantum speed-up. Additionally, \textit{quantumalgorithms.org} brings together many important quantum algorithms for data analysis \cite{qalgorg}.

\section{How Will Quantum Computing Help me with all my Data?}
\label{How}

\noindent We can see, even from the concise introduction above, that quantum computation, as it is conventionally broached, is very much bound up with the theory of computational complexity. However, when I speak to computational researchers from outside of quantum computing, invariably what they say to me is not (for example) `I am struggling with this \emph{computationally} hard problem', but rather they ask `how can quantum computing help me with all of my \emph{data}?' For we are living through an era of unprecedented data generation, and this poses problems at every stage of the computational workflow. The most urgent questions that researchers are asking of nascent computational technologies is how they can remedy these emerging and growing problems.

This is the challenge taken up by Aram Harrow in \textit{small quantum computers and large classical datasets} \cite{harrow2020small}, which proposes using the quantum computer to do computationally intensive model searches when substantial data reduction is possible on the `large classical dataset'. However, the question of what quantum computing can do to deal with `big data' more generally is tricky: loading data onto the quantum computer is well-known to be a hard problem, not only with the small-scale quantum hardware that is available at present, but a fundamental problem in principle, and one which if we are not careful could easily nullify the quantum advantage. This has in turn brought about something of a divide between `pessimists' who believe that the data-loading problem is fundamentally an insurmountable obstacle, and `optimists' who focus on the unquestionable computational benefits once the data \textit{is} loaded, and assume that some solution will emerge to the data-loading problem itself. 

The purpose of this article is to provide one answer to the motivating question of what quantum computing can do to help with the massive proliferation of data, that is neither unduly pessimistic or optimistic, but rather is \textit{realistic} -- and illuminates a plausible path ahead for the eventual integration of quantum computing into data-centric applications.

\section{Quantum Computing and Machine Learning: A Match Made in Heaven?}
\label{QC+ML}

\noindent In recent years there has been an explosion of papers on `Quantum Machine Learning' (QML), and a cynic would say that this amounts to little more than a case of buzz-word fusion to unlock funding sources and generate hype. But I am not a cynic -- for one thing some of the most respected researchers in quantum computing are working on QML -- and there are (at least) two very good reasons to believe that quantum computing may ultimately offer significant computational advantages for machine learning tasks. These two reasons in turn inform two complementary approaches to QML. 

\begin{figure*}[!t]
\begin{center}
\includegraphics[width=0.9\textwidth]{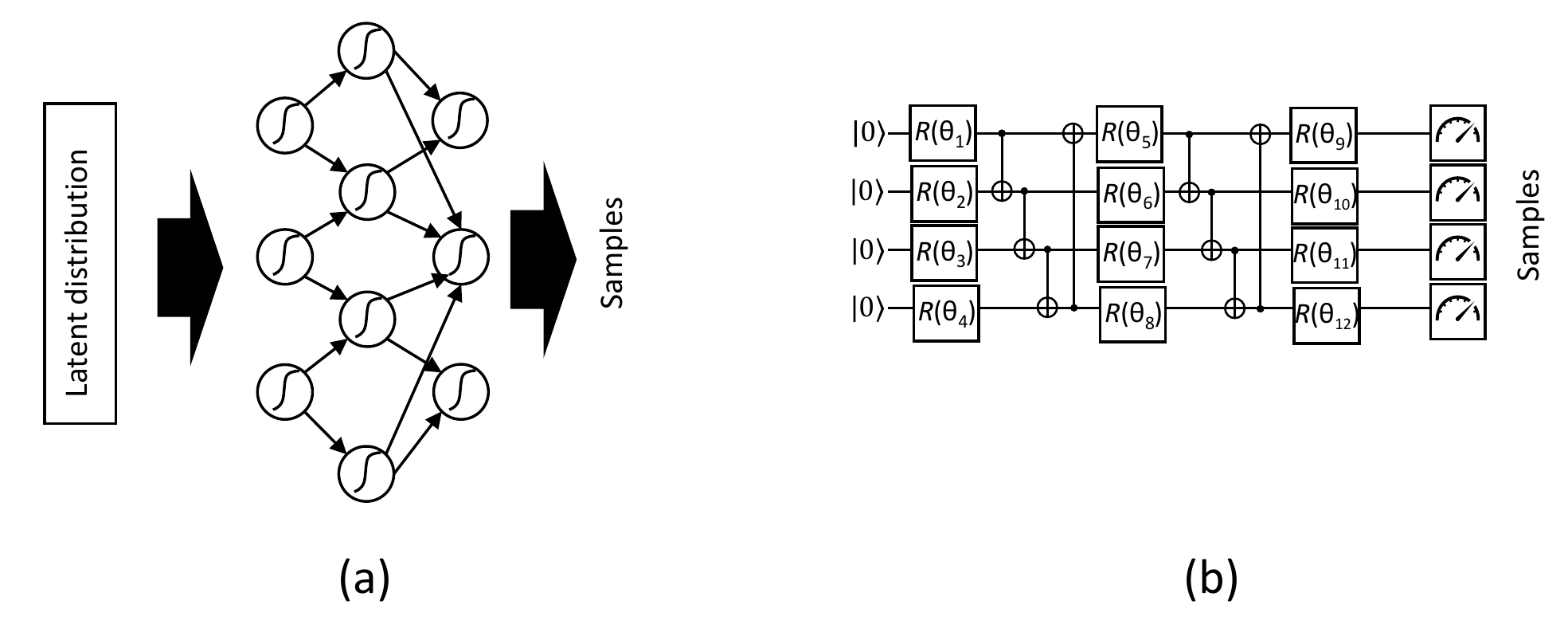}
\end{center}
\captionsetup{width=.9\linewidth}
\caption{Classical and Quantum Generative models: (a) A classical generative model will typically use an artificial neural network to map samples from some standard latent space distribution (such as a uniform or Gaussian distribution) to samples from the target distribution; (b) using a parameterised quantum circuit as a quantum generative model is a similar concept, except that the measurement corresponds to a random collapse of the quantum state, and hence suffices to generate samples from a probability distribution even if the PQC operates on a fixed initial state such as $\ket{\mathbf{0}}$}
\label{f2}
\end{figure*}

One approach stems from the functional similarity between artificial neural networks (ANNs) and parameterised quantum circuits (PQCs) \cite{Benedetti_2019b}, as shown in Fig.~\ref{f2}. In particular, by virtue of the fact that we must always measure the quantum state to extract some information (and noting that measurement triggers a probabilistic `collapse' of the quantum superposition into one of some ensemble of possible states), a PQC is innately something we sample from, and is therefore, in a sense, analogous to an ANN trained as a generative model \cite{qGAN1, qGAN2, qGEN, Zoufal2019, Benedetti_2019a, Chang2021, Chang2022}. (There are myriad proposals to use PQCs in place of ANNs for other learning tasks, \textit{e.g.} \cite{QML1, QML2, QML3}, but considering generative models suffices to illustrate my point here.) The original hope of quantum advantage in generative modelling stemmed from the fact that there is strong theoretical evidence for the existence of probability distributions from which samples can be prepared quantumly in polynomial time, but would require exponential time classically, for example probability distributions sampled by \textit{IQP circuits} \cite{IQP}. Indeed, most proposals for and demonstrations of \textit{Quantum Supremacy} are sampling experiments \cite{supreme1, supreme2}. The ramifications for generative modelling are that, should the target distribution be some such `classically intractable' distribution to sample, then we would need an infeasibly large ANN to train a generative model thereof, but only a relatively small PQC. 

However, in practice this is perhaps a slightly over-simplistic outlook: because the datasets of interest in engineering and other typical applications will themselves have been generated by some `classical' process, and so are unlikely to have probability distributions that we expect to be hard to classically sample from. (For instance, we do not, in general, expect classical random processes such as financial time-series to exhibit the sort of correlations seen in the measurement statistics of highly-entangled quantum circuits.) To put it another way, even though PQCs have greater expressivity, it is not clear that this can be harnessed for any useful application. Compounding this apparently fundamental obstacle is the fact that PQCs are incredibly hard to train \cite{bittel2021training}, and the cost function landscape is overwhelmingly dominated by large, flat regions termed \textit{barren plateaus} \cite{McClean2018, barren2, barren3, barren4, barren5}. Nevertheless, in spite of these apparent problems, there is some evidence that QML based on PQC-training will yield useful quantum advantage in classical data science applications \cite{pqcad1, pqcad2, pqcad3}. Indeed, in spite of the question marks hanging over PQCs as ML models in terms of their trainability and expressivity, there remains hope that such models may still have greater power in terms of generalisation capability \cite{eisert22}.

So we turn to the second approach to QML, which builds on the ability, in principle, of quantum computers to perform certain linear algebra computations (exponentially) faster than the best classical counterpart, and in particular suggests that this feature can be used to enhance certain machine learning and data science tasks. One recent paper has suggested that the finding \textit{Betti numbers}, a task in topological data analysis, may be feasible on NISQ machines \cite{AkhalwayaExp} -- although the question of whether the Betti numbers for which the computation can be exponentially sped-up are practically relevant has been raised \cite{McardleStreamlined}. Other than this, it is worth noting, that whilst the training of PQCs as ML models is championed by its proponents as a naturally NISQ application, the quantum enhancement of linear algebra computations is expected to require full-scale (fault-tolerant) quantum computers

The most famous quantum algorithm for linear algebra is Harrow, Hassidim and Lloyd's algorithm for solving a linear system (ubiquitously known as `HHL') \cite{HHL}. Specifically, consider the system of linear equations:

\begin{equation}
A\mathbf{x} = \mathbf{b} 
\end{equation}
which is solved by inverting $A$ and then pre-multiplying $\mathbf{b}$ by $A^{-1}$ to find $\mathbf{x}$. Classically this computation takes time that is worse than linear in the size of $A$ (even if $A$ is sparse), whereas in certain circumstances HHL runs in time that is only poly-logarithmic in the size of $A$ -- thus giving an exponential improvement over the best classical algorithms. HHL leverages the fact that an $n$-qubit quantum circuit is nothing more than a $2^n \times 2^n$ unitary matrix and thus, in a sense, a quantum computer is simply a machine that performs exponentially big matrix multiplications. When the matrix $A$ is sparse and well-conditioned (the ratio of its largest to its smallest eigenvalues is not too big) then it is possible to construct the matrix operation $A^{-1}$ as a quantum circuit. Moreover, this $n$-qubit circuit is only polynomially deep (in $n$) and hence the entire algorithm runs in time that is poly-logarithmic in the size of the matrix, $A$ (a full complexity analysis also accounts for the fact that each attempt at inversion only succeeds with a certain probability, however the overall poly-logarithmic complexity continues to hold, even when this is included).

HHL does, however, suffer from a number of caveats, one of which is the model for access to the data: it is assumed that the quantum computer has access to $\mathbf{b}$ as some quantum state $\ket{b}$, the preparation of which is not counted in the algorithm's complexity. Indeed, one can immediately see that complexity at least linear in the size of $\mathbf{x}$ (and hence the size of $A$) would be incurred even to read $\mathbf{x}$, and so an overall poly-logarithmic complexity can only be possible if the appropriate quantum state is pre-prepared. The question of the need for a reasonable data access model is one of the problems raised by Scott Aaronson when discussing the potential for quantum advantage in machine learning applications \cite{fineprint}.

This issue was clarified and generalised by Ewin Tang, who showed that all proposed QML algorithms of this second approach can be \textit{dequantised} if a classical algorithm is given commensurate data access \cite{dequant0, dequant1, dequant2, dequant4, dequant3}. `Dequantised' means that there is no \textit{exponential} quantum advantage -- although there could still be a practically useful \textit{polynomial} advantage. That is, with Tang's results, the quantum and classical algorithms both run in time polynomial in the logarithm of the size of the linear algebra objects in question, however that polynomial may be of much higher degree for the classical algorithm -- thus the quantum algorithms may still provide a practically beneficial speed-up. This was indeed the case for Tang's original dequantisation breakthrough \cite{dequant0}, where she proposed a `quantum-inspired' classical version algorithm of the \textit{quantum recommendation system} of Kerenidis and Prakash \cite{kerenidis20178th}.

Finally, it is also pertinent to note that HHL itself has only been dequantised for low-rank matrix inversion \cite{dequant3}; \cite{dequant5}: there is still an exponential quantum advantage when the matrix to be inverted is full- (or close to full-) rank. Indeed, fast matrix inversion is still seen as being a potential `killer application' of quantum computing, and is a fertile area of research in, for example partial differential equation (PDE) solving, when a finite-difference approach can be used to turn the PDE into a system of linear equations \cite{PDE1, PDE2, PDE3, PDE4, PDE5}. 

\section{Quantum Data: The Holy Grail?}

\noindent We have seen that the drawbacks with QML do not entirely diminish its potential practical utility. However, what is in some ways even more notable is that, until now, we have solely been talking about \textit{classical} data, so we may ask: `what about \textit{quantum} data?' That is, what if we have some quantum sensing or metrology process delivering quantum states directly as training data? 

Taking the two approaches to QML in turn, when manipulating quantum rather than classical data, it is certainly more reasonable to expect that there may be some fundamental reason why a QML model may be required. In particular, even if the quantum data is immediately measured to give a classical sample, in general such a sample may exhibit `non-classical' correlations that cannot be reproduced by any reasonable-sized classical algorithm (for instance, as already noted, the correlations present in measurements of highly-entangled IQP circuits are believed to need exponentially large classical circuits to reproduce). Moreover, it has been shown that, in certain instances, barren plateaus are \textit{not} present in generative modelling of \textit{quantum} data (in this case, when the quantum state itself, not a measurement thereof, is delivered as training data to the model) \cite{nobarren1, nobarren2}.

Turning to the second approach to QML, to provide commensurate data access to compare classical and quantum algorithms Tang's dequantisation results ordain the classical algorithms with \textit{sample and query} access to the data. Suppose we have some $N$-element vector $x$, `query access' means the value $x_i$ can be extracted (for any $i$), and `sample access' means we sample a number, $i$ between 0 and $N-1$ with probability $x_i / \sum_j x_j$. If the quantum state is prepared from classical data then (as Tang asserts) it is reasonable to assume that sample and query access could be attained in about the same number of operations. If, however, the data is presented as a quantum state, then only sample access is available to a classical algorithm (sample access is obtained simply by measuring the quantum state in question). This in turn implies that, when the input is quantum data, the dequantisation results no longer necessarily hold, and the possibility of exponential quantum advantage is upheld.

\section{Monte Carlo or Bust?}

\noindent Responding by basically saying `soon there may be even more data which is quantum in nature and thus intrinsically \textit{needs} QML' is only really half an answer to the motivating question of what quantum computing can do to help processing vast datasets. For the implicit emphasis in the question was on the data we already have, and expect in the immediate future. To answer this, it is helpful to step back and ask: what is it we want \textit{from} these large datasets? Invariably, the aim will be to extract some quantities pertaining to the dataset as a whole and, moreover (even if it is not immediately thought of in these terms), such quantities will usually amount to some sort of expectation of the distribution that the data has been sampled from (or simple combinations of expectation values). For instance, obviously recognisable quantities such as the mean and higher moments are expectation values, however other quantities such as various measures of risk will be found by computing an appropriate expectation. Additionally, quantities that are usually thought of not as \textit{expectations} but rather as \textit{probabilities} such as the probability of rain in a weather forecast will actually be found by numerically integrating over a number of marginal parameters\footnote{More generally, probabilities \textit{are} expectations (over indicator functions). For instance, $p(A) = \mathbb{E}(I(x \in A))$, where $I$ is the indicator function.}. 

Such a desideratum coincides with one of the (still relatively few) fundamental computational tasks that we know admits a provable quantum advantage, namely quantum Monte Carlo integration (QMCI) \cite{MontanaroMC}. Furthermore, significant progress has been made to allow such an advantage to be realised with minimal quantum resources. 

Monte Carlo integration (MCI) is the process of numerically estimating some expectation value, 
\begin{equation}
\mathbb{E}(f(x)) = \int_x f(x) p(x) \mathrm{d} x
\end{equation}
which cannot be evaluated analytically, but where the probability distribution, $p(x)$, can be sampled from (and $f(.)$ is some function). Notably, on any digital computer (classical or quantum) the integral will actually be a sum, owing to the necessary quantisation and truncation of the support of $p(x)$, thus:
\begin{equation}
\mathbb{E}(f(x)) = \sum_x f(x) p(x) \approx \frac{1}{q} \sum_{i=1}^q f(X_i)
\end{equation}
where $X_i \sim p(x)$ are i.i.d samples. The approximate equality represents the process of MCI, and in particular the mean squared error (MSE) is $\mathcal{O}(q^{-1})$. When performing high-dimensional integrals numerically, MCI is the most efficient method. (Note that quasi-Monte Carlo \cite{quasiMC} and other non-i.i.d classical methods have better convergence in $q$, but suffer the \textit{curse of dimensionality} -- the complexity grows exponentially in the number of dimensions -- and hence are inefficient for high-dimensional integrals.)

If we break down (classical) MCI, we can see that it amounts to a very simple three-step process: sample from $p(x)$; apply the function $f(.)$; and then average over many such samples with the function applied. In QMCI, there is an analogous three-step process: first we take as an input a \textit{state preparation circuit}, $P$, which prepares a quantum state $\ket{p}$ that samples from $p(x)$ when measured in the computational basis:
\begin{equation}
\label{peqn}
\ket{p} = \sum_{x} \sqrt{p(x)} \ket{x}
\end{equation}
where, for simplicity, we assume that $p(x)$ is supported over some $N = 2^n$ points.

Secondly, a circuit, denoted $R$, is applied to $\ket{p}$ with one further qubit appended such that the following state is prepared:
\begin{equation}
R \ket{p} \! \ket{0} = \sum_{x } \sqrt{p(x)} \ket{x}  \left( \sqrt{1-f(x)} \ket{0} + \sqrt{f(x)} \ket{1} \right)
\end{equation}

The circuit $R$ thus encodes the function applied, $f(.)$ and in particular has the property that, when measured, the appended qubit has probability of being one equal to:
\begin{equation}
\sum_{x } p(x)  f(x) 
\end{equation}
according to the Born rule, which tells us to square and sum all terms in the sum where the final qubit is in state $\ket{1}$. This is exactly the value we are trying to estimate with MCI, and it turns out that the (quantum) algorithm \textit{quantum amplitude estimation} (QAE) can estimate this with MSE $\mathcal{O}(q^{-2})$, where $q$ is now the number of uses of the circuit $P$ \cite{brassard2000quantum}. Accepting (for now) that this quantity `$q$' corresponds to that in classical MCI, we can see that this represents a quadratic advantage in convergence: for a certain desired MSE, only about square root as many samples are required quantumly as would be classically.

However, QAE was not originally seen as an ideal candidate as a source of \textit{near term} quantum advantage, as it uses \textit{quantum phase estimation} \cite{QPE} an algorithm that is expected to require full-scale quantum computers. That all changed with the advent of \textit{amplitude estimation without phase estimation} \cite{Suzuki2020}, which showed how to obtain the full quadratic quantum advantage, but using a number of shallow-depth circuits and classical post-processing to estimate the expectation value. A number of other proposals have since followed in the same vein \cite{grinko2019iterative, Aaronson2020, nakaji2020faster, giurgicatiron2020low}.

Two more, complementary, breakthroughs have further fueled the hope that QMCI can be a source of near-term quantum advantage:
\begin{enumerate}
\item \textit{Noise-aware quantum amplitude estimation} \cite{naqae} takes advantage of the fact that QAE circuits have a very specific structure to handle device noise as if it were estimation uncertainty. This suggests that significantly less error-correction may be needed to achieve a useful advantage in QMCI, compared to other calculations of comparable size.
\item \textit{Quantum Monte Carlo integration: the full advantage in minimal circuit depth} \cite{fourierQMCI} shows how to decompose the Monte Carlo integral as a Fourier series such that the circuit $R$, which may in general constitute an unreasonably large contribution to the total circuit depth, can be replaced by minimally deep circuits of rotation gates (this procedure is hereafter referred to as `Fourier QMCI').
\end{enumerate}
In particular, the second of these informs us of the sorts of applications that are likely to see a (relatively) early quantum advantage. For instance, Fourier QMCI is especially advantageous for numerical integrals that can be decomposed as a product of: some $p(x)$, for which a suitable encoding can prepared by a relatively shallow state preparation circuit (i.e., as described in Eqn (\ref{peqn})); and some $f(x)$ which can be extended as a piecewise periodic function whose Fourier series can be calculated and satisfies certain smoothness conditions. Areas in which computationally intensive numerical integrations are commonplace include computational fluid dynamics and high-energy physics -- indeed, in the latter QMCI solutions have begun to be explored \cite{QMCIHEP}.

For general numerical integrals the randomnness in MCI is a device to enable efficient numerical integration -- however for most data-centric MCIs we do expect that $p(x)$ will have a more literal role as the probability distribution from which the data has been sampled. This in turn raises the question of how to a construct the state preparation circuit, $P$. 

From a theoretical point of view it is always possible to construct a suitable $P$ from the corresponding classical sampling process \cite{classamp} (this resolves the earlier question of why the number of classical samples can be compared to the number of quantum uses of the state preparation circuit -- the two uses of `$q$' that were treated as equivalent), and such a result may well ultimately find practical application. For example, ANNs trained as generative models are instances of classical sampling processes, and so such generative models can be converted into suitable circuits, $P$. However, in the near-term such quantum circuits are likely to be infeasibly deep, and so instead we should focus on applications that leverage the quantum advantage in a more direct manner. In particular, the quantum advantage is manifested in a quadratic reduction in the number of samples required to attain a certain required accuracy, and so applications where a very large number of samples are required provides a good starting point -- especially when those samples are from (relatively) simple stochastic processes.

With regards to data-centric engineering and science, one helpful way to think about which applications that QMCI will impact in the near-term is in terms of the distinction between parametric and non-parametric models. For we have already established that we must operate on some model for the generation of the observed data: in cases where the best (classical) approach is to use the dataset to fit parameters from some parameterised family of distributions then we expect the corresponding circuit $P$ to be relatively easy to construct. Conversely, if the model is non-parametric (or even something like a deep neural network that, by some, may be regarded as parametric -- just with an enormous number of parameters that do not correspond in the natural and straightforward way to the statistics of the observed data as do traditional parametric models) then there is a real risk that the circuit $P$ will be hard to construct in the near-term.

\begin{figure*}[!t]
\begin{center}
\includegraphics[width=0.8\textwidth]{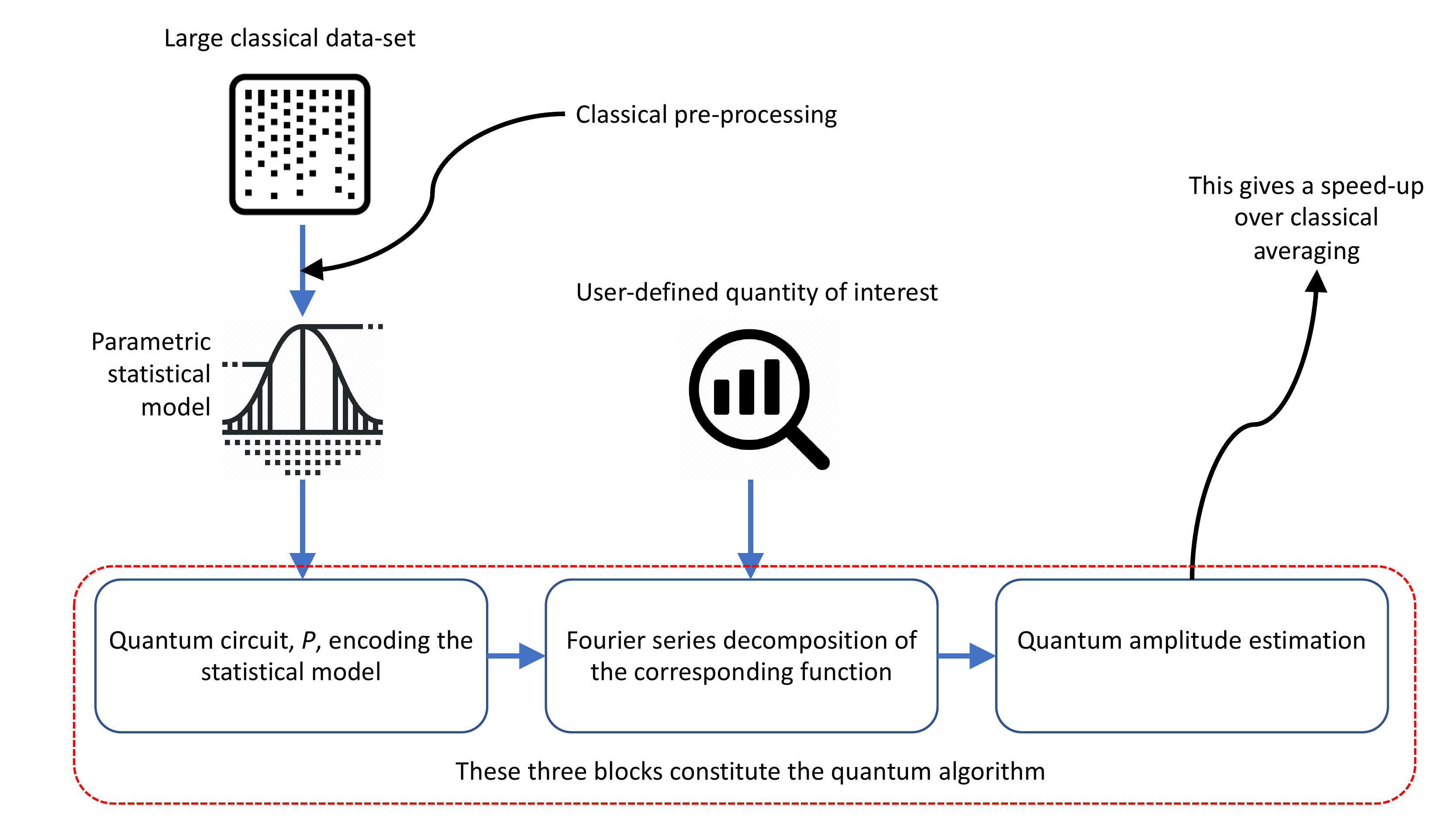}
\end{center}
\captionsetup{width=.9\linewidth}
\caption{The general framework of Fourier QMCI applied to large classical datasets. In the future (as quantum hardware matures) the parametric statistical model may be optionally replaced by a non-parametric (or ML) model; or perhaps even a quantum machine learning model. In all cases, QMCI only ever requires \textit{sample access} to the data
}
\label{f3}
\end{figure*}

Turning now to some specific examples, one promising area concerns time-series data, where some large dataset is used to fit the parameters of some model such a hidden Markov model, auto-regressive model or auto-regressive moving average model (amongst many others). Notably, using an abundance of historical data to tune the parameters of time-series models is commonplace throughout applications of MCI in financial and actuarial engineering. Indeed, it is the case that the vast majority of the early QMCI literature has focused on financial applications \cite{Rebentrost2018, Stamatopoulos_2020, Woerner2019, bouland2020prospects, QCfinance, egger2019credit, kaneko2020quantum, chakrabarti2020threshold, rebentrost2018quantum, financeppr, an2020quantumaccelerated, jpmc, goldmangradients}. 

That is not to say, however, that QMCI will ultimately only find application in financial engineering. For instance, the `function applied' in financial applications of QMCI usually corresponds to some sort of thresholded-average -- most obviously when calculating the expected return on a European option the function applied is essentially a ReLU function -- and functions of these types (i.e., piece-wise linear) are likely to be similar to suitable functions to calculate notions of `cost' in (for example) supply-chain and logistic-optimisation applications \cite{supplychain}. Moreover, Monte Carlo methods are widely used in virtually every area of data-centric engineering and science from medical imaging \cite{medimage} to chemical, biochemical and environmental \cite{biomed} to energy modelling \cite{energymod} and handling big data in general \cite{bigdata}. Indeed, rather than exhaustively cataloguing every conceivable application of QMCI to data-centric engineering and science, a better approach is to set out the general framework, as is shown in Fig.~\ref{f3}, such that expert readers can see how the quantum advantage may be realised in their respective domains. 

At first sight, the scheme laid out in Fig.~\ref{f3} may appear to side-step the central question of how quantum computing will help with large datasets, as the data handling itself represents a classical pre-processing step. However, this is simply a reflection of the reality that data-loading is generally hard and that it is prudent to focus on tasks where there is an unequivocal quantum advantage (i.e., in estimate converge as a function of number of samples). Moreover, many data-centric applications (for example those in finance) are indeed of the form where the data-loading can be achieved by fitting the parameters of some statistical model, but thereafter the statistical estimation is the bottleneck: and it is exactly those applications that quantum computing can incontrovertibly enhance.

\section{Outlook and Speculation}

\noindent In this article I have deliberately homed in on QMCI as a likely source of near-term quantum advantage in data science applications. This is a personal view, and many others still focus on the possibility that there will be genuine useful `NISQ advantage' (for example, using PQCs as ML models). However, since this \textit{is} my view, as explicitly noted in the introduction, I see the established division into the NISQ and full-scale eras of quantum computing as overly simplistic. Instead, I prefer to think about how we can use \textit{resource constrained} quantum hardware, that is neither strictly NISQ (there may be enough qubits for some mild error-correction) but neither is full-scale (in the sense that full-scale algorithms may be typified by an attitude of being unconcerned with resource demands that appear only as negligible contributions to asymptotic complexity) -- and hence bridges between the established ideas of NISQ and full-scale quantum computing. Certainly it is true that the first applications of quantum computing with a provable advantage will be those where consideration of resource management have been given special attention. For reasons that have recurred throughout this article I am of the view that QMCI shows great promise to be such an application.

In light of this, I feel it is incumbent upon me to offer some predictions about when this will come to fruition. As any responsible quantum computing researcher will attest, such predictions are hard to make at present, but now that the leading quantum hardware manufacturers are beginning to commit to roadmaps with quantified targets, it is at least possible for theorists to make loose predictions contingent on said roadmaps being met. The first comment to make is that both IBM and Google have committed to reaching 1 Million qubits by the end of the decade \cite{IBM, googlerm}, and such a figure would certainly be enough for useful quantum advantage in many applications including QMCI.

To make a more precise prediction, with regards to QMCI we benefit from the fact that a leading research group has published a paper setting out suitable benchmarks to facilitate resource estimation for when there will be useful quantum advantage in QMCI \cite{chakrabarti2020threshold}. The specific focus of the paper in question is on QMCI applications in finance, but the broad principle is likely to extend to other applications. We have estimated that our Fourier QMCI algorithm \cite{fourierQMCI} reduces resource requirements (number of quantum operations) by at least 30\% to over 90\% in some cases for the benchmarks set out, and if the circuits were re-constructed in a slightly different way we have estimated that the total number of physical superconducting qubits required would be in the 1000s to 10,000s. The exact value within this range depends on whether qubit qualities improve enough for low-overhead error-correction codes \cite{Tomita2014} to be practical; and in particular whether the overhead can be reduced by exploiting asymmetries in the noise \cite{XZZX, Fragile} -- both of which remain very active research topics.

This resource estimation places useful advantage in QMCI in the five-year horizon for the leading superconducting roadmaps and a similar timescale is likely for trapped-ion devices (eg \cite{quantinuumrm}, note that trapped-ion quantum computers typically have many fewer, but higher-quality qubits, and tend to have less specific roadmaps). This is also consistent with other predictions, for example that of \textit{QCWare} and \textit{Goldman Sachs} \cite{QCW}.

Rather than leaving the prediction at that, it is worth `playing devil's advocate' and exploring whether this is unreasonably hubristic -- particularly in light of a widely-circulated paper suggesting that near-term quantum advantage would be hard to obtain for algorithms exhibiting only a quadratic speed-up \cite{Babbush2021}. There are three central claims in \cite{Babbush2021}, which provide a useful framework to scrutinise the legitimacy of my prediction:
\begin{enumerate}
\item \textit{Quadratic-advantage quantum algorithms are dominated by circuits of Toffoli gates, which are extremely expensive to implement using error-corrected quantum computation}. This is certainly true for un-optimised algorithms, however Fourier QMCI \cite{fourierQMCI} moves to classical post-processing exactly those Toffoli-heavy circuits, whilst upholding the full quantum advantage.
\item \textit{Error-correction overheads are, in any case, expensive}. Again, this is true, which is why bespoke approaches to error-correction that exploit the specific algorithm structure and handle a certain amount of the device noise at the application level (as does noise-aware QAE \cite{naqae}) are crucial to achieve near-term quantum advantage.
\item \textit{Algorithms for which there is a quadratic quantum advantage can typically be massively parallelised when performed classically, meaning that a useful quantum advantage only occurs at much larger problems sizes, once the parallelism has been accounted for}. This is again true in the case of MCI, but leaves out one very significant detail, namely that QMCI can itself be massively parallelised. 
\end{enumerate}
The third item reveals an in important subtlety: in the (quantum computing) sector we obsess about the `route to scale' in terms of adding ever more qubits to the same chip -- but once quantum computers reach moderate scale, it will be just as important to scale up the \textit{number} of quantum computers available. Just as classical HPC can accelerate classical data-centric applications by running calculations on different cores in parallel, so it will be the case that running quantum circuits in parallel will be crucial for early advantage in data-centric applications. (To be clear, here the parallelisation is classical -- there is no need for entangled connections between the different cores -- although the subject of \textit{distributed quantum computing}, where there are entangled connections between the different quantum cores is a fascinating topic in its own right, see \textit{e.g.} \cite{dist1, dist2, dist3}). Indeed, for our above resource estimates we have only quantified when quantum hardware will be capable of running the requisite quantum circuits -- in order for this to translate into a practical benefit, sufficient quantum cores must be available to the user.

So where does that leave us? That there exist data science relevant quantum algorithms, such as QMCI, that exhibit a provable quantum advantage, coupled with the fact that the leading players are now beginning to scale up quantum hardware, provides a great cause for optimism that quantum computing will impact on data-centric engineering and science applications in the near- to medium-term. However, in quantum computing we have learnt that optimism must always go hand-in-hand with caution: there are serious engineering challenges at every layer, from the design of the quantum computer itself, to the control software, to the optimisation of the algorithms that run the desired applications. In particular, we know that preparing quantum states that encode the relevant model or distribution is usually the bottleneck in QMCI -- cracking this \textit{data-loading problem} is the key to unleashing the power of quantum computation onto myriad applications within finance, supply chain \& logistics, medical imaging and energy modelling.

\section*{Acknowledgments}
My thanks to Ewin Tang and Seth Lloyd for kindly answering my questions regarding quantum linear algebra and the dequantisation thereof -- of course, any remaining errors or misconceptions are my own. I also thank Alexandre Krajenbrink, Sam Duffield and Konstantinos Meichanetzidis for reviewing and providing valuable suggestions. Thanks also to the anonymous reviewers at DCE, whose suggestions helped to further improve the paper.

\bibliography{}

\end{document}